\documentclass[epj]{svjour}
\usepackage{graphicx,bm}
\usepackage{amssymb}
\usepackage{amsmath}

\newcommand{\up}{\uparrow}
\newcommand{\dn}{\downarrow}
\newcommand{\Eq}[1]{Eq.~\eqref{#1}}
\newcommand{\half}{{1\over 2}}
\newcommand{\midb}[1]{\left[ #1 \right]}
\newcommand{\Fig}[1]{Figure~\ref{#1}}
\newcommand{\pa}{\parallel}
\newcommand{\bigb}[1]{\left\{ #1 \right\}}
\newcommand{\Sb}{\ensuremath{\mathbb S}}
\newcommand{\RR}{\ensuremath{\mathbf R}}
\newcommand{\TT}{\ensuremath{\mathbf T}}
\newcommand{\Rb}{\ensuremath{\mathbb R}}
\newcommand{\Tb}{\ensuremath{\mathbb T}}
\newcommand{\braket}[2]{\langle #1|#2\rangle}
\newcommand{\UU}{\ensuremath{\mathbf U}}

\newcommand{\lRa}{\Longrightarrow}
\newcommand{\ra}{\rightarrow}
\newcommand{\Ra}{\Rightarrow}

\newcommand{\qwith}{\quad\mbox{with}\quad}
\newcommand{\qfor}{\quad\mbox{for}\quad}

\begin{document}

%+++++++++++++++++++++++++++++++++++++++++++++++++++++++++++++++++++++++++++
\title{A numerical method to solve the Boltzmann equation for a spin valve}

\author{Jiang Xiao\inst{1} \and A. Zangwill\inst{1} \and M. D. Stiles\inst{2}}

\institute{School of Physics, Georgia Institute of Technology, Atlanta, GA
30332-0430
\and
Electron Physics Group, National Institute of Standards and
Technology, Gaithersburg, MD 20899-8412}
\date{Received: date / Revised version: date}

\abstract{ We present a numerical algorithm to solve the Boltzmann
equation for the electron distribution function in magnetic multilayer
heterostructures with non-collinear magnetizations.  The solution is
based on a scattering matrix formalism for layers that are
translationally invariant in plane so that properties only vary
perpendicular to the planes.  Physical quantities like spin density,
spin current, and spin-transfer torque are calculated directly from
the distribution function. We illustrate our solution method with a
systematic study of the spin-transfer torque in a spin valve as a
function of its geometry.  The results agree with a
hybrid circuit theory developed by Slonczewski for geometries typical
of those measured experimentally.  }

%+++++++++++++++++++++++++++++++++++++++++++++++++++++++++++++++++++++++++++
\maketitle
%#########################################################################
%%++++++++++++++++++++++++++++++++++++++++++++++++++++++++++++
%%----------------------------------------
\section{Introduction}
\label{sec:intro}

Magnetic multilayer structures have attracted a great deal of experimental and
theoretical attention. One motivation for these studies is the potential
application of such structures for data storage and spin dependent transistors.
One special effect in a magnetic multilayer that can induce magnetic reversal or
magnetic dynamics is called spin-transfer. In the ten years since its
theoretical prediction, \cite{Slonczewski:1996,Berger:1996} spin-transfer has
been studied extensively both experimentally \cite{Katine:2000b,Urazhdin:2003a,Kiselev:2003,Rippard:2004a,Fert:2004} and
theoretically.
\cite{Sun:2000,Hernando:2000,Waintal:2000,Stiles:2002b,Li:2003,Bazaliy:2004}
One 
fundamental issue is how to reliably calculate the spin-transfer torque in spin
valve systems. To solve this problem, different approaches have been
developed, including
the Boltzmann equation \cite{Stiles:2002a,Xiao:2004}, microscopic quantum
mechanics \cite{Edwards:2005}, drift-diffusion theory 
\cite{Berger:1998,Grollier:2003,Fert:2004,Stiles:2004,Barnas:2005}, and
circuit theory \cite{Brataas:2000,Brataas:2001}. 

Each approach has its own advantages and disadvantages. Simple theories,
for example circuit theory or the drift-diffusion approach, treat the
transport in terms of densities and current densities and do not track
individual electrons.   Such methods have the advantage that they can
give analytic results in some limits and the disadvantage that they
leave out some essential physics.  One of
the major approximations of these theories is that they ignore the differences
between electrons propagating in different directions.  Fully quantum mechanical
calculations track all of the electrons and all of the coherent
scattering processes like coherent multiple scattering between layers.
However, such calculations are quite time consuming and the coherent
multiple scattering between layers that is included in such an approach does
not seem to play a role in experimental results.  

The semiclassical Boltzmann equation is a useful compromise between
these extremes. In such an approach, the scattering is treated
semiclassically.  For transport in collinear magnetic systems, it has
been used in two ways.  Superlattices that have mean free paths
longer than layer thicknesses can be treated as artificial bulk
materials \cite{Zahn:1995,Zahn:2005}.  This approach retains the
coherent multiple scattering between the interfaces.  The other
approach is to solve the Boltzmann equation within layers and join the
solutions through boundary conditions at the interface.  Here, we
describe a generalization of the latter approach to treat
non-collinear magnetizations.  In this approach, the Boltzmann
equation tracks individual electrons  
through the distribution function, but ignores the coherent 
multiple scattering.  It is easier to treat defect
scattering in such an 
approach than it is in a fully coherent calculation.  The advantage of
the Boltzmann calculation is that it is simply computable and includes
the essential physics. One disadvantage is that it cannot give
analytical results.  
The neglect of coherent multiple scattering between interfaces is both
an advantage and a disadvantage.  The greater simplicity that results
in the calculation is an advantage for the metallic devices that have been
measured to date because the coherent scattering does not appear to
play a role.  However, this neglect would be a disadvantage in
devices in which such effects were important.

Slonczewski has developed a hybrid approach combining aspects of
circuit theory with a simplified Boltzmann equation to give an analytic
expression for the spin-transfer torque
\cite{Slonczewski:2002}. This approach gives
much more accurate results than the drift-diffusion method for typical
device geometries.  However, it breaks down when layer thicknesses
become longer than are typical.

In this paper, we describe a numerical algorithm that solves the Boltzmann
equation for the electron distribution function in a magnetic multilayer system
using a scattering matrix formalism. The spin-transfer properties are calculated
from the resulting distribution function.  We compare our Boltzmann results to
the results from more approximate methods. These comparisons show
Slonczewski's hybrid  theory is highly 
accurate for typical experimental structures, but drift-diffusion shows
systematic deviations.  We have reported results of calculations
using the methods described in the present paper in earlier
papers \cite{Stiles:2002a,Xiao:2004}.

The paper is organized as follows: Section \ref{sec:gmbe} gives a
generalized spin-dependent Boltzmann equation appropriate for
ferromagnetic systems; Section \ref{sec:formal} describes in detail
the algorithm that solves the Boltzmann equation in a spin valve
magnetic multilayer system; Section \ref{sec:appl} shows the results
of the Boltzmann method developed in Section \ref{sec:formal} applied
to a spin valve. Section \ref{sec:summary} gives a summary.  Readers
who are not interested in the formalism can skip directly to
Section \ref{sec:appl}.

%%----------------------------------------
\section{Generalized matrix Boltzmann equation}
\label{sec:gmbe}

In non-magnetic materials, the spin independent Boltzmann equation is
%%<<<<<<<<<<<<<<<<<<<<
\begin{equation}
    {\bf v_k}\cdot\frac{\partial g({\bf r, k})}{\partial{\bf r}}
    -e{\bf E}\cdot{\bf v_k} {\partial g({\bf r, k})\over\partial
    \epsilon_{\bf k}}
    = \int d{\bf k'}P_{\bf k,k'}[g({\bf r, k'})-g({\bf r, k})],
    \label{eqn:Boltzmann_full}
\end{equation}
%%>>>>>>>>>>>>>>>>>>>>
where $g({\bf r,k})$ is an electron distribution function that
depends on the spatial coordinate $\bf r$ and electron wave-vector
$\bf k$.
${\bf v_k}$ is the electron velocity, $\epsilon_{\bf k}$ its energy, 
${\bf E}$ is the electric
field, and we have ignored the magnetic field.
$P_{\bf k,k'}$ is the probability of electron scattering from state ${\bf
k'}$ to state ${\bf k}$. The Boltzmann equation is valid
only when the bulk 
properties vary slowly. It cannot be used for abrupt interfaces or boundaries;
later we describe how to use boundary conditions to relate solutions
of the Boltzmann equation in different regions across the interfaces.

Transport in metals is dominated by the electrons near
the Fermi energy.  The occupancy of states far from the Fermi energy
does not change and those states do not contribute to the transport.
This suggests the use of the linearized Boltzmann equation in which
the distribution function is assumed to have the form
\begin{equation}
{g}({\bf r,k}) = f_0({\bf k}) 
+ {f}({\bf r,k}) \delta(E_{\rm F} -\epsilon_{\bf k}) ,
\end{equation}
where $f_0$ is the equilibrium distribution function.
The delta function restricts the wave vector to the Fermi surface,
$|{\bf k}|=k_{\rm F}$ for free electrons.  With this approximation for
the distribution function, the Boltzmann equation becomes
%%<<<<<<<<<<<<<<<<<<<<
\begin{equation}
    {\bf v_k}\cdot\frac{\partial f({\bf r, k})}{\partial{\bf r}}
    -e{\bf E}\cdot{\bf v_k} 
    = \int_{\rm FS} d{\bf k'}P_{\bf k,k'}[f({\bf r, k'})-f({\bf r, k})],
    \label{eqn:Boltzmann}
\end{equation}
%%>>>>>>>>>>>>>>>>>>>>
where the integration over ${\bf k'}$ is now a two-dimensional
integral restricted to the Fermi surface.  The scattering rate $P_{\bf
  k,k'}$ has been rescaled.  A delta function has been factored out of
each term.  In the second term on the left hand side, this delta
function comes from $\partial f_0 /\partial\epsilon$.

For spin dependent magnetic materials, one needs a spin dependent distribution
function as well as a spin dependent Boltzmann equation. When there is a natural
quantization axis, i.e., all electron spins are either parallel
($\sigma=\uparrow$) or anti-parallel ($\sigma=\downarrow$)
to the axis, an electron distribution function is separated into the
distribution functions for spin-up and spin-down electrons: $f^\up({\bf r, k})$
and $f^\dn({\bf r, k})$ 
%%<<<<<<<<<<<<<<<<<<<<
\begin{eqnarray}
    {\bf v_k}^\sigma\cdot{\partial f^\sigma({\bf k})\over\partial{\bf r}}
    - e{\bf E}\cdot{\bf v_k}^\sigma 
    &=& \int_{\rm FS} d{\bf k'}P_{\bf k,k'}^\sigma[f^\sigma({\bf
    k'})-f^\sigma({\bf k})]
\nonumber\\
    && + \int_{\rm FS} d{\bf k'}P_{\bf k,k'}^{\rm sf}[f^{\sigma'}({\bf k'})-f^\sigma({\bf
    k})], \nonumber\\
    \label{eqn:Boltzmann-flip}
\end{eqnarray}
%%>>>>>>>>>>>>>>>>>>>>
The $\bf r$ dependence of the distribution function has been suppressed, and
$\sigma = \up$, $\dn$ and $\sigma \neq \sigma'$. Compared with
\Eq{eqn:Boltzmann}, \Eq{eqn:Boltzmann-flip} has an additional spin flip
scattering term on the right hand side because the distribution function in
\Eq{eqn:Boltzmann} includes both spin types, while the one in
\Eq{eqn:Boltzmann-flip} is for only one spin type. Similar to the definition of
$P_{\bf k,k'}$ in \Eq{eqn:Boltzmann}, $P_{\bf k,k'}^\sigma$ and $P_{\bf
k,k'}^{\rm sf}$ are the probabilities of electron scattering from state ${\bf
k'}$ to state ${\bf k}$ without and with spin flip. We assume that $P_{\bf
k,k'}^{\rm sf}$ is the same for spin flip in both directions, up to down or down
to up.

In a non-magnet, there is no natural quantization axis, so it is
convenient to use the axis of neighboring ferromagnetic layers if the
magnetizations of those layers are collinear.  In this case, the
scattering $P_{\bf k,k'}$ is spin-independent, so the
substitutions $f^0 = \half(f^\up + f^\dn)$ and $f^{z} =
\half(f^\up - f^\dn)$ give the pair of equations
\begin{eqnarray}
    {\bf v_k}\cdot{\partial f^0({\bf k})\over\partial{\bf r}}
    & - e{\bf E}\cdot{\bf v_k}
    = & 
      \int_{\rm FS} d{\bf k'}P_{\bf k,k'}[f^0({\bf k'})-f^0({\bf k})]
\nonumber\\
    &&+ \int_{\rm FS} d{\bf k'}P_{\bf k,k'}^{\rm sf}[f^{0}({\bf k'})-f^0({\bf
    k})],
\nonumber\\
    {\bf v_k}\cdot{\partial f^z({\bf k})\over\partial{\bf r}}
    & \hskip 0.6in = &
      \int_{\rm FS} d{\bf k'}P_{\bf k,k'}[f^z({\bf k'})-f^z({\bf k})]
\nonumber\\
    &&- \int_{\rm FS} d{\bf k'}P_{\bf k,k'}^{\rm sf}[f^{z}({\bf k'})+f^z({\bf
    k})],  
\nonumber\\
    \label{eqn:Boltzmann-nmflip}
\end{eqnarray}
In the first equation, spin flip scattering acts as another form of
non-flip scattering as far as the number accumulation is concerned.
In the second equation, the electric field plays no role because it
does not couple to the spin accumulation.  In the systems of interest
here, the magnetizations are not collinear, so the spin axis in the
distribution function can vary with both ${\bf r}$ and ${\bf k}$.
However, the generalization is straightforward.  The second equation in
\Eq{eqn:Boltzmann-nmflip} is replicated for each of the other
directions in spin space, with $z\rightarrow x,\ y$.  The
generalization can be derived from a matrix form of the Boltzmann
equation in terms of the matrix distribution function
\begin{eqnarray}
    \hat{f}({\bf r,k}) 
    &=& f^0\sigma_0 + f^x\sigma_x + f^y\sigma_y + f^z\sigma_z.
\nonumber\\
&=& \hat{U}({\bf r,k}) \midb{ \begin{array}[c]{cc}
    f^\up({\bf r,k}) & 0 \\ 0 & f^\dn({\bf r,k})
    \end{array} } \hat{U}^\dag({\bf r,k})
    \label{eqn:df-pauli}
\end{eqnarray}
where  $\sigma_x, \sigma_y$ and
$\sigma_z$ are  Pauli spin matrices,  $\sigma_0$ is a $2\times 2$
identity matrix, and $\hat{U}$ is a unitary rotation matrix.
The rotation matrix allows spin direction to point in an arbitrary
direction over the Fermi surface and as a function of position.  

In strong ferromagnets, the
rotation matrix in \Eq{eqn:df-pauli} is independent of the
position on the Fermi surface or ${\bf k}$ as in
Eq.~(\ref{eqn:Boltzmann-flip}).  This constraint is a consequence of
the large difference in the Fermi surfaces for majority and
minority electrons in strong ferromagnets.  The constraint arises because
any transverse spin accumulation in the electrons near the Fermi
surface rapidly dissipates.  The transverse spins precess in the large
exchange field and do so at different rates because of the complicated
Fermi surfaces.  The precessing transverse components rapidly dephase
with respect to each other, leaving no net transverse moment.

%%----------------------------------------
\section{Formal technique}
\label{sec:formal}

%%<<<<<<<<<<<<<<<<<<<<
\begin{figure}
    \centering
    \resizebox{0.9\columnwidth}{!}{%
    \includegraphics{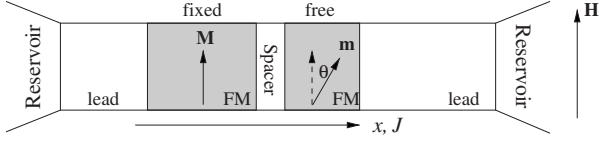}}
    \caption{Schematic view of a perpendicular spin valve structure.}
    \label{fig:spinvalve}
\end{figure}
%%>>>>>>>>>>>>>>>>>>>>

\Fig{fig:spinvalve} shows a schematic picture of a spin valve --- a magnetic
multilayer structure where a thin-film non-magnet (spacer) is sandwiched between
two thin-film ferromagnet (FM) layers. The latter connect to reservoirs with
non-magnetic leads. The main purpose of this paper is to solve the
Boltzmann equation numerically for the distribution function in such a
spin valve 
structure.  In reality, the cross sectional dimensions of these
structures are much larger than typical layer thicknesses so that the
transport in the interior of the sample is more important than that
near the edges.  The reservoirs
have much larger cross sectional areas and hence smaller resistances than the
active structures so that the transport is largely perpendicular to the
layers.  This combination of features suggests a simple model in which
the layers are treated as
translationally invariant in plane so that properties only vary
perpendicular to the planes.
Electrons move in all three directions, but the net current and all
variation is in the $x$-direction, i.e. $\hat{f}({\bf
  r,k})=\hat{f}(x,{\bf k}) $.  

The calculation proceeds in eight steps: (\ref{sec:discrete})
discretize the spin dependent Boltzmann equation using a numerical
mesh of the Fermi surface; (\ref{sec:bulksolve}) solve the discretized
Boltzmann equation for the eigensolutions in the non-magnetic and
ferromagnetic bulk; (\ref{sec:layerscatt}) use the eigensolutions to
construct the layer scattering matrix for each bulk layer in the spin
valve; (\ref{sec:intscatt}) construct the interface scattering matrix
for each interface in the spin valve; (\ref{sec:sysscat}) connect the
bulk and interface scattering matrices into a single system-wide
scattering matrix; (\ref{sec:bc}) determine the boundary conditions
and apply them to the system-wide scattering matrix to calculate the
distribution function expansion coefficients; (\ref{sec:sysdist})
calculate the distribution function values within the spin valve; and
(\ref{sec:transprop}) calculate the spin density (spin accumulation),
spin current, and spin-transfer torque using the distribution
function.
%{\it Give a outline of the method.}

%%----------------------------------------
\subsection{Discretization of the Boltzmann equation}
\label{sec:discrete}

To discretize the Boltzmann equation, we need a numerical mesh for the electron
wave-vector $\bf k$ that can accurately and simultaneously describe all Fermi
surfaces for both spin types and for all layers.  
A simple method for choosing a mesh is as follows. Choose a uniform mesh in the
direction parallel to the interfaces (perpendicular to $x$): ${\bf k}_\pa^j$.
For each material, there could be several points on the Fermi surface that have
the same ${\bf k}_\pa^j$. We label their longitudinal wave-vectors $k_x^n$.  A
complete mesh for the Fermi surface is ${\bf k}_i = (k_x^n, {\bf k}_\pa^j)$.
The mesh weights are determined by the area on each Fermi surface associated
with $k_x^n$. This mesh may converge slowly, so it may be necessary to refine
the ${\bf k}_\pa^j$ mesh to include a higher density of points. Let us assume
one such mesh has sampling points $\bigb{{\bf k}_i}_1^N$ with weighting factor
$\bigb{w_i}_1^N$. Using this mesh, the integration for any continuous function
$h({\bf k})$ on the Fermi surface can be discretized as:
%%<<<<<<<<<<<<<<<<<<<<
\begin{equation}
    \int_{\rm FS} h({\bf k}) d{\bf k}
    = \sum_{i=1}^N w_i h({\bf k}_i).
    \label{eqn:int-FS}
\end{equation}
%%>>>>>>>>>>>>>>>>>>>>

We discretize the integrations in \Eq{eqn:Boltzmann-flip} using this mesh.
Assuming the system is one dimensional in $x$ direction, we have a discretized
form of the Boltzmann equation:
\begin{equation}
    {\partial f_i^\sigma\over\partial x} - e E_x
    = \sum_{j,\sigma'} [\hat{V}^{-1}\hat{B}]_{ij}^{\sigma\sigma'} f_j^{\sigma'}, 
    \label{eqn:Boltzmann-disc}
\end{equation}
%>>>>>>>>>>>>>>>>>>>>
where the subscript $i$ and $j$ mean that $\bf k$ is evaluated at ${\bf k}_i$ or
${\bf k}_j$, for instance $f_i^\sigma = f^\sigma({\bf k}_i)$. The velocity
matrix $\hat{V}$ and the scattering matrix $\hat{B}$ are $2N\times 2N$ matrices
($N$ from ${\bf k}_i$ index, $2$ from spin index) with matrix elements
%%<<<<<<<<<<<<<<<<<<<<
\begin{eqnarray}
    V_{ij}^{\sigma\sigma'}
    &=& v_i^{\sigma x}\delta_{ij}^{\sigma\sigma'}, \\
    B_{ij}^{\sigma\sigma'}
    &=&  w_j P_{ij}^\sigma\delta_{\sigma\sigma'}
    - {\delta_{ij}^{\sigma\sigma'}\over\tau_i^\sigma}
    + w_j P_{ij}^{\rm sf}(1-\delta_{\sigma\sigma'})
    - {\delta_{ij}^{\sigma\sigma'}\over\tau_i^{\rm sf}},
\nonumber
    \label{eqn:Bij}
\end{eqnarray}
%%>>>>>>>>>>>>>>>>>>>>
where 
$\delta_{ij}^{\sigma\sigma'} = \delta_{ij}
\delta_{\sigma\sigma'}$,  
$1/\tau_i^\sigma=\sum_j w_j P_{ij}^\sigma$, is the spin-dependent
non-spin-flip scattering rate, and
$1/\tau_i^{\rm sf}=\sum_j w_j
P_{ij}^{\rm sf}$ is the spin flip scattering rate.  See
Eq.~(\ref{eqn:Boltzmann-flip}). 
The matrix $\hat{B}$ is singular because
$\sum_{j,\sigma'}B_{ij}^{\sigma\sigma'} = 0$.

%%----------------------------------------
\subsection{Solution of the Boltzmann equation in the bulk}
\label{sec:bulksolve}

In this section, we briefly summarize the results of
Ref.~\cite{Penn:1999} and then generalize them to treat systems with
non-collinear magnetizations.
According to Ref.~\cite{Penn:1999}, the Boltzmann equation
\Eq{eqn:Boltzmann-disc} can be solved by breaking it into a particular
equation  
%%<<<<<<<<<<<<<<<<<<<<
\begin{equation}
    \sum_{j,\sigma'}~[\hat{V}^{-1}\hat{B}]_{ij}^{\sigma\sigma'}
    f_j^{\sigma'} = - e E_x,
    \label{eqn:Boltzmann-par}
\end{equation}
%%>>>>>>>>>>>>>>>>>>>>
and a homogeneous equation,
%%<<<<<<<<<<<<<<<<<<<<
\begin{equation}
    \sum_{j,\sigma'}\bigb{
     \delta_{ij}^{\sigma\sigma'} {\partial\over\partial x}
    - [\hat{V}^{-1}\hat{B}]_{ij}^{\sigma\sigma'}}
    f_j^{\sigma'} = 0.
    \label{eqn:Boltzmann-homo}
\end{equation}
%>>>>>>>>>>>>>>>>>>>>
The solution to the particular equation \Eq{eqn:Boltzmann-par} is
%%<<<<<<<<<<<<<<<<<<<<
\begin{equation}
    F^\sigma_0(x,{\bf k}_i) = -e E_x \sum_{j,\sigma'} 
    [\hat{B}^{-1}\hat{V}]_{ij}^{\sigma\sigma'}.
    \label{eqn:sol-par}
\end{equation}
%>>>>>>>>>>>>>>>>>>>>>
Due to the singularity of $\hat{B}$, $\hat{B}^{-1}\hat{V}$ is not the inverse of
$\hat{V}^{-1}\hat{B}$. Instead, the inverse matrix is defined as in Section 2.9
of Ref.~\cite{Press:1986} using a singular value decomposition.  Since
the constant 
vector is perpendicular to the null space, the solution is well
defined.  Physically, the solution is the distribution function of the
current in bulk material.  Since this distribution is well defined
physically, we expect it to be mathematically as well, if we have
formulated the problem correctly.

Most solutions to the homogeneous equation \Eq{eqn:Boltzmann-homo} vary
exponentially in space:
%%<<<<<<<<<<<<<<<<<<<<
\begin{equation}
    F^\sigma_n(x,{\bf k}_i) = g^\sigma_n({\bf k}_i)e^{\lambda_n x}
    \qwith n\in[3,2N],
    \label{eqn:sol-homo-ex}
\end{equation}
%>>>>>>>>>>>>>>>>>>>>
%%%~~~~~~~~~~~~~~~~~~~~
where $g_n^\sigma({\bf k}_i)=g^{i\sigma}_n$ and $\lambda_n$ are the $n$-th
eigenvector and eigenvalue of $\hat{V}^{-1}\hat{B}$
%%<<<<<<<<<<<<<<<<<<<<
\begin{equation}
    \sum_{j,\sigma'}~[\hat{V}^{-1}\hat{B}]_{i\sigma,j\sigma'}
    g_n^{j\sigma'} = \lambda_n g_n^{i\sigma}.
\end{equation}
%%>>>>>>>>>>>>>>>>>>>>
See the Appendix in Ref.~\cite{Penn:1999} for more details.
Since the bulk is translationally invariant in $x$ direction, half of the
eigenvalues $\lambda_n$ are positive and half are negative.
The matrix $\hat{V}^{-1}\hat{B}$ is defective, which means that the
degenerate zero eigenvalue only has one eigenvector.  Because it is
defective, the homogeneous equation has two solutions that do not
have exponential form
%%<<<<<<<<<<<<<<<<<<<<
\begin{eqnarray}
    F^\sigma_1(x,{\bf k}_i) &=& 1 \nonumber\\
    F^\sigma_2(x,{\bf k}_i)  &=& x F^\sigma_1(x,{\bf k}_i) + 
    \sum_{j,\sigma'} [\hat{B}^{-1}\hat{V}]_{ij}^{\sigma\sigma'}.
    \label{eqn:sol-homo}
\end{eqnarray}
%>>>>>>>>>>>>>>>>>>>>
which can be verified by plugging them back in
\Eq{eqn:Boltzmann-homo}.  

Physically, $F_1$ is the solution describing a uniform shift of the chemical
potential and $F_2$ describes a current carrying solution having a spatially
varying density and a associated uniform diffusion current.  The rest
of eigensolutions, \Eq{eqn:sol-par} and \Eq{eqn:sol-homo-ex}, are
exponential solutions that are necessary near interfaces.  These
solutions are not allowed in a uniform bulk because they diverge in
one direction.  Together, these solutions describe arbitrary solutions
of the Boltzmann equation in each layer. 

Both the particular solution $F_0$ and the homogeneous solution $F_2$
are current carrying because of the ${\bf k}_i$-dependence of their
summation terms.  But, they describe current associated with
different processes. $F_0$ describes the current due to the electric
field, and $F_2$ describes the current due to density
gradients. Formally, $F_2 = A (E_x x - F_0)$, where $A$ is a constant,
so $F_0$ plus the electric field $E_x$ can be interchanged with $F_2$.
Computationally, since we work with uniform current, we solve
everything with $E_x = 0$ and no $F_0$ and work with $F_2$. Once we find the
coefficient of $F_2$, the physical solution is the corresponding $F_0$
and $E_x$. 
That is, we solve the problem as if the uniform current was
exclusively due to diffusion with no electric field and then
reinterpret the charge accumulation as an electric potential and set
the charge accumulation to zero.  This interpretation makes sense due
to the short screening length in metals.

There is a natural quantization axis in a ferromagnet defined by its
magnetization. The distribution function is described by the
eigensolutions in \Eq{eqn:sol-homo}: 
%%<<<<<<<<<<<<<<<<<<<<
\begin{eqnarray}
    \hat{f}(x,{\bf k}_i) &=& f^\up \sigma_\up + f^\dn \sigma_\dn
    \nonumber\\
    f^\sigma &=& \sum_{n=1}^{2N} \alpha_n^\sigma F^\sigma_n(x,{\bf k}_i)
    \equiv \vec{F}^\sigma\cdot\vec{\alpha^\sigma}.
    \nonumber\\
    \label{eqn:FM-sol}
\end{eqnarray}
%%>>>>>>>>>>>>>>>>>>>>
Here, $\sigma_\up=(1/2)({\rm I}+\sigma_z)$, $\sigma_\dn=(1/2)({\rm I}-\sigma_z)$,
and the $\alpha_n^\sigma$ are the expansion coefficients. 

In a non-magnet, there is no natural quantization axis, so
we write the distribution function as in
\Eq{eqn:df-pauli}. We construct a different basis set $F_n^{\tau}$,
$\tau=0,x,y,z$ from 
$F_n^{\up,\dn}$ for the non-magnet. First, we know that the eigenvectors
$F_n^\sigma(x,{\bf k}_i)$ in \Eq{eqn:sol-homo-ex} and \Eq{eqn:sol-homo} break
into separate eigenvectors for charge transport with $F_n^\up = F_n^\dn$ and
spin transport with $F_n^\up=-F_n^\dn$. For instance, the eigenvectors with
$n=1$ and 2 correspond to charge transport because $F_{1,2}^\up = F_{1,2}^\dn$.
In general, half of the eigenvectors (assume for the first half: $n\in[1,N]$)
corresponds to the charge transport; the other half (for the second half:
$n\in[N+1,2N]$) is for the spin transport. Therefore, the basis set in a
non-magnet can be constructed as:
%%<<<<<<<<<<<<<<<<<<<<
\begin{subequations}
\begin{align}
    F^{\tau=0}_n(x,{\bf k}_i) 
    &= \half\midb{F^\up_n(x,{\bf k}_i) + F^\dn_n(x,{\bf k}_i)}
    & n\in [1,N],\\
    F^{\tau=x,y,z}_n(x,{\bf k}_i) 
    &= \half\midb{F^\up_{n+N}(x,{\bf k}_i) - F^\dn_{n+N}(x,{\bf k}_i)}
    & n\in [1,N].
\end{align}
\label{eqn:NM-basis}
\end{subequations}
%%>>>>>>>>>>>>>>>>>>>>
Thus, in the non-magnet layers, the general solution for the distribution
function is 
%%<<<<<<<<<<<<<<<<<<<<
\begin{eqnarray}
    \hat{f}(x,{\bf k}_i) &=& f^0\sigma_0 + f^x\sigma_x + f^y\sigma_x + f^z\sigma_z
    \nonumber\\
    f^s  &=& \sum_{n=1}^N \alpha_n^s F_n^s(x,{\bf k}_i)
    \equiv \vec{F}^s\cdot\vec{\alpha}^s,
    \label{eqn:NM-sol}
\end{eqnarray}
%%>>>>>>>>>>>>>>>>>>>>
where $s = 0, x, y, z$, and $\alpha_n^s$ are expansion coefficients.
\Eq{eqn:NM-basis} tells us that $f^x$, $f^y$, and $f^z$ share the same set of
eigenvectors. This is because the Boltzmann solution does not depend on the
choice of spin quantization axis in a non-magnet. 

%%----------------------------------------
\subsection{Layer scattering matrix}
\label{sec:layerscatt}

%%<<<<<<<<<<<<<<<<<<<<
\begin{figure*}
    \centering
    \resizebox{1.5\columnwidth}{!}{%
    \includegraphics{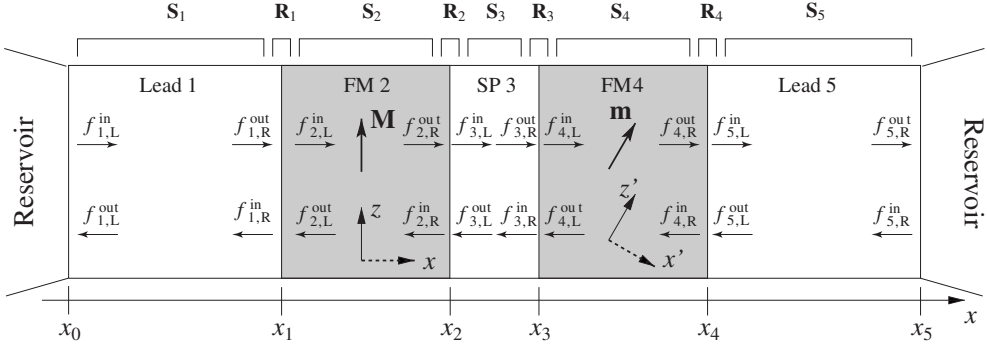}}
    \caption{Scattering matrices.  The center of the figure lists the
    incoming and outgoing distribution functions at each of the
    layers.  The top lists the layer, $S_n$, and interface, $R_n$
    scattering matrices that relate the distribution functions.}
    \label{fig:scat-matrix}
\end{figure*}
%%>>>>>>>>>>>>>>>>>>>>

The eigensolutions are used to construct a scattering matrix which relates the
boundary values of each layer. \Fig{fig:scat-matrix} shows a schematic picture
of a spin valve, where lead/FM/spacer/FM/lead are labeled layer 1 to layer 5,
and $x_{0,1,2,3,4,5}$ denotes the $x$ coordinate of each interface. In
\Eq{eqn:FM-sol} and \Eq{eqn:NM-sol}, we expand the distribution function in each
layer using a set of eigensolutions. From these expansions we construct the
(layer) scattering matrix for layer m, $\Sb_m$, which relates the incoming
electron distribution functions $f_{\rm m, L/R}^{\rm in}$ and the outgoing ones
$f_{\rm m, L/R}^{\rm out}$ at the left (L) and right (R) sides of the layer m
(\Fig{fig:scat-matrix}):
%%<<<<<<<<<<<<<<<<<<<<
\begin{equation}
    \midb{\begin{array}{c}
	f_{\rm m,L}^{\rm out} \\
	f_{\rm m,R}^{\rm out}
    \end{array}}
    = \Sb_m
    \midb{\begin{array}{c}
	f_{\rm m,L}^{\rm in} \\
	f_{\rm m,R}^{\rm in}
    \end{array}}.
    \label{eqn:layer-sm}
\end{equation}
%%>>>>>>>>>>>>>>>>>>>>
Here, the matrices in the ferromagnet ($m=$ 2, 4) and non-magnet($m=$
1, 3, 5) are respectively
%%<<<<<<<<<<<<<<<<<<<<
\begin{eqnarray}
        \left[ f_{\rm m,L}^{\rm in} \right]^{\rm T} &=&
        \left[ f_{\rm m,L}^{\rm in, \up}, 
        f_{\rm m,L}^{\rm in, \dn} \right]
\nonumber\\
        \left[ f_{\rm m,L}^{\rm in} \right]^{\rm T} &=&
        \left[ f_{\rm m,L}^{\rm in, 0} ,
        f_{\rm m,L}^{\rm in, x} ,
        f_{\rm m,L}^{\rm in, y} ,
        f_{\rm m,L}^{\rm in, z} \right] .
    \label{eqn:ud0xyz}
\end{eqnarray}
%%>>>>>>>>>>>>>>>>>>>>
and $f_{\rm m, L}^{\rm in, \up} = f_{\rm m}^\up(x_{m-1}, {\bf k}_i^+)$, and
$f_{\rm m, R}^{\rm in, \up} = f_{\rm m}^\up(x_m, {\bf k}_i^-)$, etc. ${\bf
k}_i^\pm$ is the electrons'
wave-vector on the Fermi surface with the (+/-) superscript indicating
whether the electorn is right (+) or left (-) going. The superscripts
``in'' and ``out'' denote 
electrons going into or out of the layer.  The definition in \Eq{eqn:ud0xyz}
generalizes in a straightforward manner for ``R'' and ``out.''

From \Eq{eqn:FM-sol}, we can express the distribution function as $f_{\rm
m}^\sigma(x, {\bf k}_i) = \vec{F}^\sigma(x, {\bf k}_i)\cdot\vec{\alpha}^s$ in
the ferromagnetic layers (m = 2, 4), then
%%<<<<<<<<<<<<<<<<<<<<
\begin{eqnarray}
    \midb{\begin{array}{c}
	f_{\rm m,L}^{\rm out} \\
	f_{\rm m,R}^{\rm out}
    \end{array}}
    &=& 
    \midb{\begin{array}{c}
	\vec{F}^\up(x_{m-1}, {\bf k}_i^-) \\
	\vec{F}^\dn(x_{m-1}, {\bf k}_i^-) \\
	\vec{F}^\up(x_m, {\bf k}_i^+) \\
	\vec{F}^\dn(x_m, {\bf k}_i^+) 
    \end{array}} \cdot\vec{\alpha}
    \equiv \vec{\bf F}_m^{\rm out}\cdot\vec{\alpha}
\nonumber\\
    \midb{\begin{array}{c}
	f_{\rm m,L}^{\rm in} \\
	f_{\rm m,R}^{\rm in}
    \end{array}}
    &=&
    \midb{\begin{array}{c}
	\vec{F}^\up(x_{m-1}, {\bf k}_i^+) \\
	\vec{F}^\dn(x_{m-1}, {\bf k}_i^+) \\
	\vec{F}^\up(x_m, {\bf k}_i^-) \\
	\vec{F}^\dn(x_m, {\bf k}_i^-) 
    \end{array}} \cdot\vec{\alpha}
    \equiv \vec{\bf F}_m^{\rm in}\cdot\vec{\alpha},
\end{eqnarray}
%%>>>>>>>>>>>>>>>>>>>>
where $\vec{\bf F}_m^{\rm in}$ and $\vec{\bf F}_m^{\rm out}$ are both $2N\times
2N$ square matrices. Therefore, \Eq{eqn:layer-sm} implies that the layer
scattering matrix for ferromagnetic layers (m = 2, 4) is
%%<<<<<<<<<<<<<<<<<<<<
\begin{equation}
    \Sb_m = 
    \vec{\bf F}_m^{\rm out} \cdot \midb{\vec{\bf F}_m^{\rm in}}^{-1}.
\end{equation}
%%>>>>>>>>>>>>>>>>>>>>
%{\it From the expansion, relate the outgoing distribution function values at the
%bulk boundary to the incoming distribution function values at the bulk boundary.
%The relating matrix is the layer scattering matrix.}
The layer scattering matrix for non-magnetic layers (m = 1, 3, 5) is constructed
in a similar way using \Eq{eqn:NM-sol} as the expansion and \Eq{eqn:ud0xyz} for
m = 1, 3, 5.

%%----------------------------------------
\subsection{Interface scattering matrix}
\label{sec:intscatt}

Using the layer scattering matrix, we are able to relate the distribution
function values at the two sides of a bulk layer. Next, we find an
interface scattering matrix which connects the distribution functions across an
interface. Right at the interface, the Boltzmann equation is not valid because the
material properties vary rapidly. The distribution functions
across the interface are related through the scattering matrix for the electron
wave-functions at the interface.

Like the layer scattering matrix in Eq.~(\ref{eqn:layer-sm}), the
interface scattering matrix $\Rb$ relates the incoming to the outgoing
distribution functions.  If the interface is specular, when electrons
scatter  from the interface, the component of the wave vector that is
parallel to the interface is conserved.  In this 
case, the interface scattering matrix is block diagonal, with non-zero
elements only for those states with the same parallel wave vector.  On
the other hand, defect scattering at the interface couples states with
different parallel wave vectors and the interface scattering matrix
becomes dense.  In the present work, we neglect defect scattering at
the interfaces for simplicity.  Interfacial defect scattering has been
considered by several authors \cite{Dugaev:1995,Zhang:1996,Xia:2006}.
Xia et al.~\cite{Xia:2006} treated interdiffusion at Co/Cu interfaces
with first principles calculations and found that such defects caused
only minor changes in the average transport properties for these
interfaces. 

Consider an isolated interface between a non-magnet and a
ferromagnet (NM/FM)  and
choose the spin quantization axis to be parallel to the magnetization of the
ferromagnet.  The interface between Lead 1 and FM2 in \Fig{fig:scat-matrix} is
such an interface if those two materials were extended to
infinity. Suppose the wave-function for an electron on the Fermi
surface in 
the corresponding NM is $\phi({\bf k}_i)$, which is orthonormal to other states
on the Fermi surface: $\braket{\phi({\bf k}_i)}{\phi({\bf k}_j)} = \delta_{ij}$.
The wave-function for an electron on the Fermi surface in the FM is
$\psi_\sigma({\bf k}_i)$, which is also orthonormal to other states on the Fermi
surface: $\braket{\psi_\sigma({\bf k}_i)}{\psi_{\sigma'}({\bf k}_j)} =
\delta_{ij} \delta_{\sigma\sigma'}$. 

In the non-magnet, the wave-function for an electron moving toward the interface
with spin pointing in an arbitrary direction is written as a linear combination
of spin-up and spin-down components:
%%<<<<<<<<<<<<<<<<<<<<
\begin{equation}
    \Phi_{\rm in} = 
    \midb{ \begin{array}[c]{c}
	a \phi({\bf k}_i^+) \\ b \phi({\bf k}_i^+)
    \end{array} },
\end{equation}
%%>>>>>>>>>>>>>>>>>>>>
where $a$ and $b$ are the coefficients of the up and down spinor components.
This incident state is scattered at the NM/FM interface, and the scattered
states are
%%<<<<<<<<<<<<<<<<<<<<
\begin{eqnarray}
    \Phi_{\rm ref} &=& \sum_j 
    \midb{ \begin{array}[c]{c}
	a R^{\rm NN,\up}_{ij} \phi({\bf k}_j^-) \\ 
	b R^{\rm NN,\dn}_{ij} \phi({\bf k}_j^-) 
    \end{array} }
    \qfor x < 0,
    \nonumber\\
    \Phi_{\rm tr} &=& \sum_j  
    \midb{ \begin{array}[c]{c}
	a T^{\rm NF,\up}_{ij} \psi_\up({\bf k}_j^+) \\ 
	b T^{\rm NF,\dn}_{ij} \psi_\dn({\bf k}_j^+) 
    \end{array} }
    \qfor x > 0,
\end{eqnarray}
%%>>>>>>>>>>>>>>>>>>>>
where $R^{\rm NN,\sigma}_{ij}$ and $T^{\rm NF,\sigma}_{ij}$ are the reflection
and transmission amplitudes for electron from ${\bf k}_i$ to ${\bf k}_j$ for
spin-up and spin-down electrons: $\sigma = \up, \dn$.

The $2\times 2$ matrix distribution function is then defined by the outer
product of the spinor coefficients. For instance, for the incident state
%%<<<<<<<<<<<<<<<<<<<<
\begin{equation}
    \hat{f}_{\rm in} = 
    \midb{ \begin{array}[c]{cc}
	aa^* & ab^* \\
	ba^* & bb^*
    \end{array} }.
\end{equation}
%%>>>>>>>>>>>>>>>>>>>>
%{\it Express the distribution function in terms of wave-functions.}
%then we have
Straightforward algebra reveals
%%<<<<<<<<<<<<<<<<<<<<
\begin{eqnarray}
    \hat{f}_{\rm ref}({\bf k}_j^-) &=& 
    {\RR_{ij}^{\rm NN}}^\dag\hat{f}_{\rm in}({\bf k}_i^+)\RR_{ij}^{\rm NN}
    \nonumber\\
    \hat{f}_{\rm tr}({\bf k}_j^+) &=& 
    {\TT_{ij}^{\rm NF}}^\dag\hat{f}_{\rm in}({\bf k}_i^+)\TT_{ij}^{\rm NF},
    \label{eqn:fRT}
\end{eqnarray}
%%>>>>>>>>>>>>>>>>>>>>
where the reflection matrix (NM to NM) and transmission matrix (NM to FM) are
%%<<<<<<<<<<<<<<<<<<<<
\begin{eqnarray}
    \RR_{ij}^{\rm NN} &=& \midb{ \begin{array}[c]{cc}
	R^{\rm NN, \up}_{ij} & 0 \\
	0 & R^{\rm NN, \dn}_{ij}
    \end{array} }
    \nonumber\\
    \TT_{ij}^{\rm NF} &=& \midb{ \begin{array}[c]{cc}
	T^{\rm NF, \up}_{ij} & 0 \\
	0 & T^{\rm NF, \dn}_{ij}
    \end{array} }.
    \label{eqn:RRTT}
\end{eqnarray}
%%>>>>>>>>>>>>>>>>>>>>

%%<<<<<<<<<<<<<<<<<<<<
%\begin{figure}
%    \centering
%    \resizebox{0.9\columnwidth}{!}{%
%    \includegraphics{be_fig3.eps}}
%    \caption{Distribution functions near the interface.}
%    \label{fig:interface-f}
%\end{figure}
%%>>>>>>>>>>>>>>>>>>>>

Considering the electrons incident onto the interface from both sides of the
interface, the scattering relationship \Eq{eqn:fRT} becomes
%%<<<<<<<<<<<<<<<<<<<<
\begin{subequations}
    \begin{align}
    \hat{f}_{\rm N}({\bf k}_j^-) &= 
    \sum_i {\RR_{ij}^{\rm NN}}^\dag\hat{f}_{\rm N}({\bf k}_i^+)\RR_{ij}^{\rm NN} +
    \sum_i {\TT_{ij}^{\rm FN}}^\dag\hat{f}_{\rm F}({\bf k}_i^-)\TT_{ij}^{\rm FN}, \\
    \hat{f}_{\rm F}({\bf k}_j^+) &= 
    \sum_i {\RR_{ij}^{\rm FF}}^\dag\hat{f}_{\rm F}({\bf k}_i^-)\RR_{ij}^{\rm FF} +
    \sum_i {\TT_{ij}^{\rm NF}}^\dag\hat{f}_{\rm N}({\bf k}_i^+)\TT_{ij}^{\rm NF}.
    \end{align}
    \label{eqn:inter-scat-matrix}
\end{subequations}
%%>>>>>>>>>>>>>>>>>>>>
%{\it Scattering coefficients for wave-function leads to scattering coefficients
%for distribution function.}

The matrix forms of the distribution functions $\hat{f}$ in
\Eq{eqn:inter-scat-matrix} are expanded using $\sigma_{\up,\dn}$ in
the FM or Pauli matrices $\sigma_{0, x, y, z}$ in the NM as in 
\Eq{eqn:df-pauli}. We represent the distribution functions by their expansion
components $f_{\up,\dn}$ in the FM and $f_{0,x,y,z}$ in the NM as in
\Eq{eqn:ud0xyz}. After this transformation, the scattering formula
\Eq{eqn:inter-scat-matrix} is written as
%%<<<<<<<<<<<<<<<<<<<<
\begin{equation}
    \midb{\begin{array}{c}
	f_{\rm N}^{\rm in} \\
	f_{\rm F}^{\rm in}
    \end{array}}
    \equiv \midb{\begin{array}{c}
    f_{\rm N}^0(x_0, {\bf k}_i^-) \\ 
    f_{\rm N}^x(x_0, {\bf k}_i^-) \\ 
    f_{\rm N}^y(x_0, {\bf k}_i^-) \\ 
    f_{\rm N}^z(x_0, {\bf k}_i^-) \\
    f_{\rm F}^\up(x_0, {\bf k}_i^+) \\ 
    f_{\rm F}^\dn(x_0, {\bf k}_i^+)
    \end{array}}
    = \Rb
    \midb{\begin{array}{c}
    f_{\rm N}^0(x_0, {\bf k}_i^+) \\ 
    f_{\rm N}^x(x_0, {\bf k}_i^+) \\ 
    f_{\rm N}^y(x_0, {\bf k}_i^+) \\ 
    f_{\rm N}^z(x_0, {\bf k}_i^+) \\
    f_{\rm F}^\up(x_0, {\bf k}_i^-) \\ 
    f_{\rm F}^\dn(x_0, {\bf k}_i^-)
    \end{array}}
    \equiv \Rb
    \midb{\begin{array}{c}
	f_{\rm N}^{\rm out} \\
	f_{\rm F}^{\rm out}
    \end{array}},
    \label{eqn:scattering-NF}
\end{equation}
%%>>>>>>>>>>>>>>>>>>>>
where $\Rb$ is the interface scattering matrix for the distribution functions
across the interface, and the matrix elements in $\Rb$ are obtained from the
scattering matrices $\RR$ and $\TT$ in \Eq{eqn:RRTT}.  Note the
apparent reversal of ``in'' and ``out'' in
Eq.~(\ref{eqn:scattering-NF}) compared to 
that in  in Eq.~(\ref{eqn:layer-sm}).  The directions in and out are
defined with respect to the layers, rather than the interfaces and the
incoming distribution for one of the layers is the outgoing
distribution for one of the interfaces.

%%----------------------------------------
\subsection{System scattering matrix}
\label{sec:sysscat}

With all the layer and interface scattering matrices $\Sb_{1,2,3,4,5}$ and
$\Rb_{1,2,3,4}$ (see \Fig{fig:scat-matrix}), we are ready to construct a
system-wide scattering matrix ${\mathbb T}$ that relates the incoming and
outgoing distribution functions near the left and right reservoirs. The
scattering matrix ${\mathbb T}$ is obtained by joining all the layer scattering
matrices and interface scattering matrices in order: $\Tb =
\Sb_1-\Rb_1-\Sb_2-\Rb_2-\Sb_3-\Rb_3-\Sb_4-\Rb_4-\Sb_5$.

The joining procedure is as follows. First we join $\Sb_1$ to $\Rb_1$ (see
\Fig{fig:scat-matrix}), where $\Sb_1$ is the layer scattering matrix that covers
the interval $[x_0^+, x_1^-]$ (layer 1), and $\Rb_1$ is the interface scattering
matrix that covers $[x_1^-, x_1^+$] (the interface at $x_1$):
%%<<<<<<<<<<<<<<<<<<<<
\begin{eqnarray}
    \midb{\begin{array}{c}
    f_{\rm 1,L}^{\rm out} \\
    f_{\rm 1,R}^{\rm out}
    \end{array}}
    &=& \midb{\begin{array}{cc}
    \Sb_1^{\rm LL} & \Sb_1^{\rm LR} \\
    \Sb_1^{\rm RL} & \Sb_1^{\rm Rb}
    \end{array}}
    \midb{\begin{array}{c}
    f_{\rm 1,L}^{\rm in} \\
    f_{\rm 1,R}^{\rm in}
    \end{array}}
    \nonumber\\
    \midb{\begin{array}{c}
    f_{\rm 1,R}^{\rm in} \\
    f_{\rm 2,L}^{\rm in}
    \end{array}}
    &=& \midb{\begin{array}{cc}
    \Rb_1^{\rm LL} & \Rb_1^{\rm LR} \\
    \Rb_1^{\rm RL} & \Rb_1^{\rm Rb}
    \end{array}}
    \midb{\begin{array}{c}
    f_{\rm 1,R}^{\rm out} \\
    f_{\rm 2,L}^{\rm out}
    \end{array}}.
    \label{eqn:SR-1}
\end{eqnarray}
%%>>>>>>>>>>>>>>>>>>>>
Here we have subdivided the distribution vectors (discretized
functions) into two subvectors for the values on the left (L) and
right (R).  The $\Sb$ and $\Rb$ matrices are correspondingly
subdivided into four submatrices labelled  LL, LR, RL, and RR, where,
for example, the LR submatrix connects incoming values on the right to
outgoing values on the left.
We denote
the joint scattering matrix by $\Tb_{\rm li}$, where
the subscript ``l'' denotes a lead, ``i'' denotes an
NM/FM or FM/NM interface, ``f'' denotes a ferromagnet, and ``n'' denotes the
non-magnetic spacer layer.
$\Tb_{\rm li}$ covers $[x_0^+, x_1^+]$ and relates $f_{\rm 1, L}^{\rm
  in/out}$ and $f_{\rm 2, L}^{\rm in/out}$:
%%<<<<<<<<<<<<<<<<<<<<
\begin{equation}
    \midb{\begin{array}{c}
    f_{\rm 1,L}^{\rm out} \\
    f_{\rm 2,L}^{\rm in}
    \end{array}}
    = \Tb_{\rm li}
    \midb{\begin{array}{c}
    f_{\rm 1,L}^{\rm in} \\
    f_{\rm 2,L}^{\rm out}
    \end{array}}.
\end{equation}
%%>>>>>>>>>>>>>>>>>>>>
By eliminating the intermediate distribution functions $f_{\rm 1,R}^{\rm
in/out}$ in \Eq{eqn:SR-1}, we have
%%<<<<<<<<<<<<<<<<<<<<
\begin{equation}
    \Tb_{\rm li} =
    \midb{\begin{array}{cc}
    \Tb^{\rm LL} & \Tb^{\rm LR} \\
    \Tb^{\rm RL} & \Tb^{\rm RR} 
    \end{array}},
    \label{eqn:join}
\end{equation}
where
\begin{eqnarray}
    \Tb^{\rm LL} &=&\Sb_1^{\rm LL}
    + \Sb_1^{\rm LR}\Rb_1^{\rm LL}(1-\Sb_1^{\rm RR}\Rb_1^{\rm LL})^{-1}\Sb_1^{\rm RL} 
\nonumber\\    
    \Tb^{\rm LR} &=& \Sb_1^{\rm LR}
    [1+\Rb_1^{\rm LL}(1-\Sb_1^{\rm RR}\Rb_1^{\rm LL})^{-1}\Sb_1^{\rm
    RR}]\Rb_1^{\rm LR} 
\nonumber\\
    \Tb^{\rm RL} &=& \Rb_1^{\rm RL}(1-\Sb_1^{\rm RR}\Rb_1^{\rm LL})^{-1}\Sb_1^{\rm RL} 
\nonumber\\
    \Tb^{\rm RR} &=&\Rb_1^{\rm RR}
    + \Rb_1^{\rm RL}(1-\Sb_1^{\rm RR}\Rb_1^{\rm LL})^{-1}\Sb_1^{\rm
    RR}\Rb_1^{\rm LR} .
\end{eqnarray}
%%>>>>>>>>>>>>>>>>>>>>
The scattering matrices described here and the method of joining them
is more complicated than approaches using transfer matrices, which
relate the boundary values from one side to the other rather than
outgoing to incoming boundary values.  However, transfer matrix
approaches become unstable as layers become thick and the present
method does not.

Using the same procedure as above we can join $\Tb_{\rm li}$ with $\Sb_2$ and
then with $\Rb_2$ to have a scattering matrix $\Tb_{\rm lifi}$ which covers
$[x_0^+, x_2^+]$. 
We continue to construct $\Tb_{\rm lif}$, $\Tb_{\rm lifi}$, $\Tb_{\rm
lifin}$, $\Tb_{\rm lifini}$, $\Tb_{\rm lifinif}$, $\Tb_{\rm lifinifi}$, and
$\Tb_{\rm lifinifil}$.
When coming to the the spacer layer, the eigensolutions and
the scattering matrix $\Sb_3$ use $z'$-axis instead of the $z$-axis (which is
used for $\Tb_{\rm lifi}$) as the spin quantization axis (see
\Fig{fig:scat-matrix}). Therefore, we have to make an axis rotation at the
spacer layer when joining $\Tb_{\rm lifi}$ with $\Sb_3$.  

The distribution functions that $\Tb_{\rm lifi}$ and $\Sb_3$ relate are:
%%<<<<<<<<<<<<<<<<<<<<
\begin{eqnarray}
    \midb{\begin{array}{c}
    f_{\rm 1,L}^{\rm out} \\
    f_{\rm 3,L}^{\rm in}
    \end{array}}
    &=& \Tb_{\rm lifi}
    \midb{\begin{array}{c}
    f_{\rm 1,L}^{\rm in} \\
    f_{\rm 3,L}^{\rm out}
    \end{array}}
    \nonumber\\
    \midb{\begin{array}{c}
    f_{\rm 3,L}^{\rm out'} \\
    f_{\rm 3,R}^{\rm out'}
    \end{array}}
    &=& \Sb_3
    \midb{\begin{array}{c}
    f_{\rm 3,L}^{\rm in'} \\
    f_{\rm 3,R}^{\rm in'}
    \end{array}},
    \label{eqn:Sb}
\end{eqnarray}
%%>>>>>>>>>>>>>>>>>>>>
The primed distribution functions are written using the $z'$-axis as the spin
quantization axis. To join $\Tb_{\rm lifi}$ with $\Sb_3$, we write $f_{\rm
3,L}^{\rm in/out}$ in terms of $f_{\rm 3,L}^{\rm in/out'}$: $f_{\rm 3,L}^{\rm
in/out}=\hat\UU^\dag f_{\rm 3,L}^{\rm in/out'}$. $\hat\UU$ is the component
representation [\Eq{eqn:ud0xyz}] of the unitary rotation matrix $\hat{U}$ in
\Eq{eqn:df-pauli}: $\hat{f}' = \hat{U}\hat{f}\hat{U}^\dag \Ra f' = \hat\UU f$.
Then 
%%<<<<<<<<<<<<<<<<<<<<
\begin{equation}
    \midb{\begin{array}{c}
    f_{\rm 1,L}^{\rm out} \\ \hat\UU^\dag f_{\rm 3,L}^{\rm in'}
    \end{array}}
    = \Tb_{\rm lifi}
    \midb{\begin{array}{c}
    f_{\rm 1,L}^{\rm in} \\ \hat\UU^\dag f_{\rm 3,L}^{\rm out'}
    \end{array}}
    \lRa
    \midb{\begin{array}{c}
    f_{\rm 1,L}^{\rm out} \\ f_{\rm 3,L}^{\rm in'}
    \end{array}}
    = 
     \Tb_{\rm lifi}'
    \midb{\begin{array}{c}
    f_{\rm 1,L}^{\rm in} \\ f_{\rm 3,L}^{\rm out'}
    \end{array}},
\end{equation}
%%>>>>>>>>>>>>>>>>>>>>
with the rotated scattering matrix
%%<<<<<<<<<<<<<<<<<<<<
\begin{equation}
    \Tb_{\rm lifi}' =
    \midb{\begin{array}{cc}
    1 & 0 \\ 0 & \hat\UU
    \end{array}}
    \Tb_{\rm lifi}
    \midb{\begin{array}{cc}
    1 & 0 \\ 0 & \hat\UU^\dag
    \end{array}}.
\end{equation}
%%>>>>>>>>>>>>>>>>>>>>
$\Tb_{\rm lifi}'$ is joined with $\Sb_3$ using the same procedure as in
\Eq{eqn:join} and becomes $\Tb_{\rm lifin}$. Continuing by joining
$\Tb_{\rm lifin}$ 
to $\Rb_3, \Sb_4, \Rb_4$ and $\Sb_5$ gives a system wide scattering
matrix $\Tb=\Tb_{\rm lifinifil}$ that relates the distribution functions near
the left reservoir and the right reservoir:
%%<<<<<<<<<<<<<<<<<<<<
\begin{equation}
    \midb{\begin{array}{c}
    f_1(x_0, {\bf k}_i^-) \\
    f_5'(x_5, {\bf k}_i^+)
    \end{array}}
    \equiv
    \midb{\begin{array}{c}
    f_{\rm 1,L}^{\rm out} \\
    f_{\rm 5,R}^{\rm out'}
    \end{array}}
    = \Tb
    \midb{\begin{array}{c}
    f_{\rm 1,L}^{\rm in} \\
    f_{\rm 5,R}^{\rm in'}
    \end{array}}
    \equiv \Tb
    \midb{\begin{array}{c}
    f_1(x_0, {\bf k}_i^+) \\
    f_5'(x_5, {\bf k}_i^-)
    \end{array}}.
    \label{eqn:system-sc}
\end{equation}
%%>>>>>>>>>>>>>>>>>>>>
\Eq{eqn:system-sc} is a condition on the boundary values of the distribution
functions for an arbitrary solution of the Boltzmann equation in the multilayer. 

In the case that the two leads (layer 1 and layer 5) are semi-infinite, we need
a system scattering matrix $\Tb$ that covers the interval $[x_1^-, x_4^+]$:
%%<<<<<<<<<<<<<<<<<<<<
\begin{equation}
    \midb{\begin{array}{c}
    f_1(x_1, {\bf k}_i^-) \\
    f_5'(x_4, {\bf k}_i^+)
    \end{array}}
    = \Tb
    \midb{\begin{array}{c}
    f_1(x_1, {\bf k}_i^+) \\
    f_5'(x_4, {\bf k}_i^-)
    \end{array}}.
    \label{eqn:system-sc2}
\end{equation}
%%>>>>>>>>>>>>>>>>>>>>

\Eq{eqn:system-sc} and \Eq{eqn:system-sc2} each have a total of $8N$ unknown
coefficients in the expansions of $f_1$ and $f_5$ but only $4N$ equations. To
solve this problem, we study the boundary conditions, which apply restrictions
on the distribution functions $f_1$ and $f_5$, and so reduce the number of
unknowns in the expansions.

%%----------------------------------------
\subsection{Boundary conditions}
\label{sec:bc}

By examining the properties of the leads and reservoirs, we can restrict the
form of the distribution functions $f_1$ and $f_5$ in the leads, such that the
number of unknown coefficients in the expansions equals the number of equations
in \Eq{eqn:system-sc} or \Eq{eqn:system-sc2}. Thus, we can uniquely determine
the distribution functions from \Eq{eqn:system-sc} and \Eq{eqn:system-sc2}. We
treat semi-infinite leads and finite leads differently, and separate the
discussion into two cases: 1) the leads are semi-infinite: the left/right lead
extends to the left/right infinitely, and 2) both leads have finite
length before connecting to reservoirs. 

\subsubsection{Semi-infinite leads}
\label{sec:infleads}

If the leads are semi-infinite, the distribution function $f_1$ in the left
lead (layer 1) includes only the exponential eigensolutions with $\lambda_n >
0$. Then from \Eq{eqn:NM-sol}, we have
%%<<<<<<<<<<<<<<<<<<<<
\begin{subequations}
\begin{align}
    f_1^0(x, {\bf k}_i) &= 
    \alpha_1^0 F_1^0(x,{\bf k}_i) + F_2^0(x,{\bf k}_i)
    + \mathop{\sum_{n=3}^N}_{\lambda_n>0} \alpha_n^0
    F_n^0(x,{\bf k}_i), \\
    f_1^{x,y,z}(x, {\bf k}_i) 
    &= \mathop{\sum_{n=1}^{N}}_{\lambda_{n+N}>0} 
    \alpha_n^{x,y,z} F_n^{x,y,z}(x,{\bf k}_i).
\end{align}
\label{eqn:f1}
\end{subequations}
%%>>>>>>>>>>>>>>>>>>>>
%%%<<<<<<<<<<<<<<<<<<<<
%\begin{equation}
%    f_1^s(x, {\bf k}_i) = 
%    \alpha_1^s F_1^s(x, {\bf k}_i) + \alpha_2^s F_2^s(x,{\bf k}_i) 
%    + \mathop{\sum_{n=3}^{2N}}_{\lambda_n>0} \alpha_n^s g_n^s({\bf k}_i) e^{\lambda_n x},
%\label{eqn:f1}
%\end{equation}
%%%>>>>>>>>>>>>>>>>>>>>
We set $\alpha_2^0 = 1$ to fix the current, because $F_2$ is the current
carrying term. Due to the $x$-translational invariance of the exponential
eigensolutions, half of the eigenvalues $\lambda_n$ are positive and half are
negative for charge transport and spin transport, respectively. Therefore there
are $2N$ unknown $\alpha_n^{0,x,y,z}$ in total, $N/2$ for each component of
$0, x, y$, and $z$.

Similarly, the distribution function $f_5'$ in the right lead (layer 5) includes
only the exponential eigensolutions with $\lambda_n < 0$:
%%<<<<<<<<<<<<<<<<<<<<
\begin{subequations}
\begin{align}
    {f_5'}^0(x, {\bf k}_i) &= 
    \beta_2^0 F_2^0(x,{\bf k}_i)
    + \mathop{\sum_{n=3}^N}_{\lambda_n<0} \beta_n^0 F_n^0(x,{\bf k}_i), \\
    {f_5'}^{\tau}(x, {\bf k}_i) 
    &= \mathop{\sum_{n=1}^{N}}_{\lambda_{n+N}<0} 
    \beta_n^{\tau} F_n^{\tau}(x,{\bf k}_i).
\end{align}
\label{eqn:f5}
\end{subequations}
%%>>>>>>>>>>>>>>>>>>>>
where $\tau=x,y,z$.
The choice $\beta_1^0 = 0$ sets the potential at the right lead to be zero.
\Eq{eqn:f5} also has $2N$ unknowns, $\beta_n^{\tau}$, $\tau=0,x,y,z$, in total.

Plugging \Eq{eqn:f1} and \Eq{eqn:f5} into \Eq{eqn:system-sc2}, we have $4N$
equations with $4N$ unknowns, the expansion coefficients $\alpha_n$'s and
$\beta_n$'s can be solved. This gives the distribution functions in the left and
right leads. The distribution function values inside the spin valve are
calculated from the boundary values using the appropriate scattering matrices,
as will be shown later.

\subsubsection{Finite leads}
\label{sec:finite_leads}

In actual spin valve samples, the leads are usually short, after which
the sample connects to essentially bulk material, here referred to as
reservoirs.  Electrons that 
enter the reservoir are much less likely to scatter back into the
sample than to stay in the reservoir until they are thermalized.  This
means that the reservoirs behave as perfect absorbers.  The
distribution of the electrons coming out of the reservoir is
characteristic of the bulk, independent of the distribution of
electrons coming in.  The connection between the reservoirs and leads
has been studied in more detail by Berger \cite{Berger:2004} and
Hamerle et al. \cite{Hamrle:2005}. 
When the leads have finite length and are connected to electron reservoirs, the
exponential eigensolutions have no singularities and all of them should be
included. In such a case, the form of the distribution functions is constrained
by the following properties of a reservoir: the electrons leaving the reservoir
have a bulk-like distribution function and the electrons with arbitrary
distribution function can be absorbed by the reservoir.
%%~~~~~~~~~~~~~~~~~~~~
%%~~~~~~~~~~~~~~~~~~~~
%{\it For reservoirs, the boundary condition is set by the property of
%reservoirs, where the distribution function for incoming electrons is shifted
%bulk distribution function, the distribution function for outgoing electrons is
%arbitrary. This boundary condition also reduces the number of unknown
%coefficients to the number of equations in the scattering matrix.}
Based on these two properties of reservoirs, we propose that the distribution
function near the reservoirs should satisfy: (1) for the electrons going from
the reservoir to the lead, the distribution function is bulk-like, namely
$f_{\rm 1,L}^{\rm in}$ and $f_{\rm 5,R}^{\rm in'}$ are spin independent and have
contributions only from $F_1^0$ and $F_2^0$ in \Eq{eqn:NM-basis}; (2) for the
electrons going from the lead to the reservoir, the distribution function has
whatever structure it wishes, namely $f_{\rm 1,L}^{\rm out}$ and $f_{\rm
5,R}^{\rm out'}$ have contributions from all $F_n^{0,x,y,z}$.

To determine the form of the distribution functions near the reservoirs, let us
first use $F_p^{0,x,y,z}$ to denote the eigensolutions with positive
eigenvalues: $\lambda_p > 0$; and use $F_q^{0,x,y,z}$ to denote the
eigensolutions with negative eigenvalues: $\lambda_q < 0$. We construct a new
set of basis functions $G_n^{0,x,y,z}$ using linear combinations of the old
basis functions $F_n^{0,x,y,z}$ such that,
%%<<<<<<<<<<<<<<<<<<<<
\begin{subequations}
\begin{align}
    G_{1,2}^{0,x,y,z} &= F_{1,2}^{0,x,y,z}, \\
    G_p^{0,x,y,z}(x_0,{\bf k}_i^+) &= 0, \\
    G_q^{0,x,y,z}(x_5,{\bf k}_i^-) &= 0. 
\end{align}
\label{eqn:new-bs}
\end{subequations}
%%>>>>>>>>>>>>>>>>>>>>
Essentially, $G_p$ is obtained by using the linear combination of $F_1(x_0, {\bf
k}_i^+)$ and $F_q(x_0,{\bf k}_i^+)$ to cancel $F_p(x_0,{\bf k}_i^+)$, and
similarly $G_q$ is obtained by using $F_1(x_5, {\bf k}_i^-)$ and $F_p(x_5,{\bf
k}_i^-)$ to cancel $F_q(x_5,{\bf k}_i^-)$.

$G_{1,2}$ and $G_p$ form the basis for the electrons that have bulk-like
right-going behavior at $x = x_0$, and $G_{1,2}$ and $G_q$ form the basis for
electrons that have bulk-like left-going behavior at $x = x_5$. The distribution
functions that satisfy the requirements (1) and (2) in the leads are constructed
as the following:
%%<<<<<<<<<<<<<<<<<<<<
\begin{subequations}
\begin{align}
    f_1^0(x,{\bf k}_i)
    &= \alpha_1^0 G^0_1(x,{\bf k}_i) + G^0_2(x,{\bf k}_i) + \sum_p \alpha_p^0 G^0_p(x,{\bf k}_i), \\
    f_1^{x,y,z}(x,{\bf k}_i)
    &= \sum_p \alpha_p^{x,y,z} G^{x,y,z}_p(x,{\bf k}_i), \\
    {f'}_5^0(x,{\bf k}_i)
    &= \beta_2^0 G^0_2(x,{\bf k}_i) + \sum_q \beta_q^0 G^0_q(x,{\bf k}_i), \\
    {f'}_5^{x,y,z}(x,{\bf k}_i)
    &= \sum_q \beta_q^{x,y,z} G^{x,y,z}_q(x,{\bf k}_i).
\end{align}
\label{eqn:f1f5}
\end{subequations}
%%>>>>>>>>>>>>>>>>>>>>
In these equations, $\alpha_2^0=1$ fixes the current, and $\beta_1^0=0$ fixes
the chemical potential at the right boundary to be zero. Since the indexes $p$
and $q$ each take $(N-2)/2$ values for the 0-component and $N/2$
values for the
$\{x,y,z\}$-components, the total number of unknown coefficients in \Eq{eqn:f1f5}
is $4N$. Therefore, by plugging \Eq{eqn:f1f5} into \Eq{eqn:system-sc}, the
coefficients $\alpha_n^s, \beta_n^s$ can be determined.
%{\it Relating the distribution function in new eigensolutions using system-wide
%scattering matrix, then solve for the unknown coefficients.}

%%----------------------------------------
\subsection{System distribution function}
\label{sec:sysdist}

%%<<<<<<<<<<<<<<<<<<<<
\begin{figure}
    \centering
    \resizebox{0.9\columnwidth}{!}{%
    \includegraphics{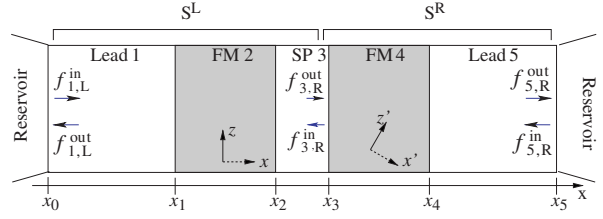}}
    \caption{Back-propagation matrices.}
    \label{fig:back-sc}
\end{figure}
%%>>>>>>>>>>>>>>>>>>>>

Once we have the distribution function values at the boundaries, either near the
lead/FM interface for the semi-infinite lead case or near the reservoir for the
finite lead case, we can calculate the distribution function values everywhere
inside the spin valve using the scattering matrices. For example, assume the
scattering matrices $\Sb^L$ covers the interval $[x_0^+,x_3^-]$ and $\Sb^R$
covers the interval $[x_3^-, x_5^-]$:
%%<<<<<<<<<<<<<<<<<<<<
\begin{eqnarray}
    \midb{\begin{array}{c}
    f_{\rm 1,L}^{\rm out} \\
    f_{\rm 3,R}^{\rm out'}
    \end{array}}
    &=& \midb{\begin{array}{cc}
    \Sb^L_{\rm LL} & \Sb^L_{\rm LR} \\
    \Sb^L_{\rm RL} & \Sb^L_{\rm RR}
    \end{array}}
    \midb{\begin{array}{c}
    f_{\rm 1,L}^{\rm in} \\
    f_{\rm 3,R}^{\rm in'}
    \end{array}}
    \nonumber\\
    \midb{\begin{array}{c}
    f_{\rm 3,R}^{\rm in'} \\
    f_{\rm 5,R}^{\rm out'}
    \end{array}}
    &=& \midb{\begin{array}{cc}
    \Sb^R_{\rm LL} & \Sb^R_{\rm LR} \\
    \Sb^R_{\rm RL} & \Sb^R_{\rm RR}
    \end{array}}
    \midb{\begin{array}{c}
    f_{\rm 3,R}^{\rm out'} \\
    f_{\rm 5,R}^{\rm in'}
    \end{array}}.
    \label{eqn:back-3R}
\end{eqnarray}
%%>>>>>>>>>>>>>>>>>>>>
%{\it Construct intermediate scattering matrix from reservoir to anywhere in the spin
%valve.}
$f_{\rm 3, R}^{\rm in/out'}$ can be solved from \Eq{eqn:back-3R}, but the
equations in \Eq{eqn:back-3R} are redundant, so we choose the half of the
equations that use the incoming boundary values rather than the outgoing ones,
i.e.,
%%<<<<<<<<<<<<<<<<<<<<
\begin{subequations}
\begin{align}
    f_{\rm 3,R}^{\rm out'}
    &= \Sb^L_{\rm RL} f_{\rm 1,L}^{\rm in} + \Sb^L_{\rm RR} f_{3,R}^{\rm in'}, \\
    f_{\rm 3,R}^{\rm in'}
    &= \Sb^R_{\rm LL} f_{\rm 3,R}^{\rm out'} + \Sb^R_{\rm LR} f_{5,R}^{\rm in'}.
\end{align}
\end{subequations}
%%>>>>>>>>>>>>>>>>>>>>
From these equations, we can calculate $f_{\rm 3, R}^{\rm in/out'}$. Similarly,
we can calculate the distribution function value elsewhere using a different
pair of scattering matrices $\Sb^L$ and $\Sb^R$.
%{\it Using the distribution function values near the reservoirs, and the
%intermediate scattering matrix to get the distribution function vale anywhere
%in the spin valve.}

%%----------------------------------------
\subsection{Transport properties}
\label{sec:transprop}

With the distribution functions in hand, it is straightforward to calculate
transport properties $h(\bf k)$ by integrating over the whole Fermi surface:
%%<<<<<<<<<<<<<<<<<<<<
\begin{equation}
    h = \int_{\rm FS} h({\bf k}) d\vec{k}`.
\end{equation}
%%>>>>>>>>>>>>>>>>>>>>
Using \Eq{eqn:int-FS}, the integrations for spin density and spin current are
discretized as
%%<<<<<<<<<<<<<<<<<<<<
\begin{subequations}
\begin{align}
    \mbox{spin density:}\qquad
    n_m^s(x) &= \sum_{i=1}^N w_i f^s_m(x,{\bf k}_i),\\
    \mbox{spin current:}\qquad
    \vec{j}_m^s(x) &= \sum_{i=1}^N w_i {\bf v}_i f^s_m(x,{\bf k}_i),
\end{align}
\end{subequations}
%%>>>>>>>>>>>>>>>>>>>>
where $s=0,x,y,z$ for $m=1,3,5$ (non-magnetic layers), and $s=\up,\dn$ for
$m=2,4$ (ferromagnetic layers). The spin current at $x=x_3^-$ is written in
the $x'$-$z'$ frame:
%%<<<<<<<<<<<<<<<<<<<<
\begin{equation}
    {\bf Q}(x_3^-) = j_3^x(x_3^-)\hat{\bf x}' + j_3^z(x_3^-)\hat{\bf z}',
\end{equation}
%%>>>>>>>>>>>>>>>>>>>>
where the $j_3^z(x_3^-)$ is the longitudinal piece parallel to the right FM
layer's magnetization ($z'$ direction), and $j_3^x(x_3^-)$ is the piece
perpendicular to $z'$-axis. From Ref.~\cite{Stiles:2002b} we know that the
perpendicular spin current is absorbed at the NM/FM interface, therefore the
spin-transfer torque acting on the right FM layer is
%%<<<<<<<<<<<<<<<<<<<<
\begin{equation}
    {\bf N}_{\rm st} = j_3^x(x_3^-)\hat{\bf x}'.
\end{equation}
%%>>>>>>>>>>>>>>>>>>>>
%{\it Distribution function solutions give all transport properties.}

%%----------------------------------------
\section{Applications}
\label{sec:appl}

In this section, we apply our numerical method to calculate the
spin-transfer torque acting on the right ferromagnetic layer of a
model spin valve.  We compare the results with equivalent calculations
using the drift-diffusion approach and Slonczewski's hybrid
theory.  First, we describe the approximations we make to simplify the
calculation so as to focus on the differences between the different
approaches.

%%----------------------------------------
\subsection{Approximations and their rationale}
\label{sec:approx}

Compared with the drift-diffusion method and circuit theory, the most
important feature of the Boltzmann method is its treatment of
electrons in the bulk moving in different directions.  In circuit
theory, the average over the Fermi surface, the spin accumulation, is used
to characterize the electrons inside that node (here a layer).  In
this treatment all of the electrons inside a node are effectively
aligned.  In the drift diffusion approximation, the distribution is
modeled by its first moment, the spin accumulation, and second moment
with respect to velocity, the spin current.  This allows for greater
flexibility in describing the distribution function, but clearly if
higher moments are important, neglecting them will lead to errors. In
a Boltzmann equation treatment of the transport the distribution
function is allowed its full flexibility.  

To study the different
approaches we consider a simple model in which we ignore the actual
shape and/or size of the Fermi surfaces and assume that the Fermi
surfaces in both the non-magnet and the ferromagnet (both spin-up and
spin-down) are perfectly spherical and are the same size. We use different
mean free paths to distinguish the differences between the electrons
in the non-magnet and the ferromagnet: $l_{\rm N}$ for non-magnet,
$l_{\rm F}^{\up,\dn}$ for spin-up and spin-down electrons in
ferromagnet.  The choice of identical and spherical Fermi surfaces
makes finding a wave vector mesh particularly simple.  
Since the problem is azimuthally symmetric, there are no contributions to the
distribution function that are not azimuthally symmetric so that only
polar variation need be considered.  We choose Gauss-Legendre
sampling for the polar direction with typically 40 mesh points.

For the interface scattering coefficients in \Eq{eqn:RRTT}, we treat
the case of ideal interfaces with no defect scattering.
Consider an
electron with wave-vector $k_x$ incident on an interface. With the
Fermi surface assumption made in the previous paragraph, the wave-vectors for
the reflected and transmitted electrons are $-k_x$ and $k_x$, respectively. The
matrix elements of the reflection and transmission matrices $\RR$ and $\TT$ in
\Eq{eqn:RRTT} are calculated in Ref.~\cite{Stiles:2000}. To allow for finite and
spin-dependent interface resistance in the equal-Fermi-surface model, we assume
$\delta$-function scattering at the interface to give the following transmission
and reflection probabilities:
%%<<<<<<<<<<<<<<<<<<<<
\begin{subequations}
\begin{align}
    |R_{ij}^{\rm NN,\sigma}|^2 = |R_{ij}^{\rm FF,\sigma}|^2
    &= {\alpha_\sigma\over\alpha_\sigma + (k_i^x)^2}\delta_{k_i^x,-k_j^x}, \\
    |T_{ij}^{\rm NF,\sigma}|^2 = |T_{ij}^{\rm FN,\sigma}|^2
    &= {(k_i^x)^2\over\alpha_\sigma + (k_i^x)^2}\delta_{k_i^x,k_j^x},
\end{align}
\label{eqn:inter-scat}
\end{subequations}
%%>>>>>>>>>>>>>>>>>>>>
where $\sigma=\up$ or $\dn$, and $k_i^x$ is a discretization of $k_x$. The
parameter $\alpha_\sigma$ is proportional to the square root of the
strength of the $\delta$-function-like interface potential. This can be read
off from the horizontal axis in Fig. 1 of Ref.~\cite{Stiles:2000} using
experimental spin dependent interface resistance data.

We also use the relaxation-time approximation: $P_{ij}^\sigma = P^\sigma =
A_{\rm FS}/\tau_\sigma$ and $P_{ij}^{\rm sf} = P^{\rm sf} = A_{\rm FS}/\tau_{\rm
sf}$, where $A_{\rm FS}$ is the area of the Fermi surface. In this limit, the
current carrying eigensolution $F_2^\sigma$ in \Eq{eqn:sol-homo} reduces to
$F_2^\sigma(x,{\bf k}_i) = x + v_i^x l^\sigma/v_F$, where $l^\sigma$ is the mean
free path for different spins and $v_F$ is the Fermi velocity.

%%<<<<<<<<<<<<<<<<<<<<
\begin{table}
    \caption{Material parameters used in the Boltzmann calculation.}
    \label{tab:bolt-parameters}
    \begin{tabular}{clrlr}
\hline\noalign{\smallskip}
    Parameter       	& Material & Value     & Units     & Reference \\
\noalign{\smallskip}\hline\noalign{\smallskip}
    $l$    		& Cu 	   & 110       &nm     	& \cite{Stiles:2002a}\\
    $l_{\rm sf}$	& Cu 	   & 450       &nm     	& \cite{Bass:1999} \\
    $l^\up$	    	& Co 	   & 16.25     &nm     	& \cite{Stiles:2002a}\\
    $l^\dn$    		& Co 	   & 6.01      &nm     	& \cite{Stiles:2002a}\\
    $l_{\rm sf}$   	& Co 	   & 59        &nm     	& \cite{Yang:1994} \\
\noalign{\smallskip}\hline\noalign{\smallskip}
    $\alpha_\up$	& Co/Cu    & 0.051     & 	& \cite{Stiles:2002a} \\
    $\alpha_\dn$	& Co/Cu    & 0.393     & 	& \cite{Stiles:2002a} \\
\noalign{\smallskip}\hline
    \end{tabular}
\end{table}
%%>>>>>>>>>>>>>>>>>>>>

Using the algorithm described above in Sec.~\ref{sec:gmbe} and the
approximations discussed 
above (equal, spherical Fermi surfaces), we calculated the spin-transfer
torque for the spin valve shown 
in \Fig{fig:spinvalve}. For the rest of this paper, we assume the
non-magnetic and ferromagnetic layers are composed of Cu and Co,
respectively. The results are quite similar if Cu and Co are replaced
by other non-magnetic and ferromagnetic metal. The input values in the
Boltzmann calculation are listed in Table \ref{tab:bolt-parameters}.

%%----------------------------------------
\subsection{Results and Comparisons}

In this section we test the drift-diffusion approach and Slonczewski's
hybrid theory by comparing the results with those found with
the Boltzmann equation.
The spin-transfer torque acting on the right ferromagnet layer ($\bf
m$ layer in \Fig{fig:spinvalve}) in a spin valve can generally be
written in the following form \cite{Slonczewski:1996}:
%%<<<<<<<<<<<<<<<<<<<<
\begin{equation}
    {\bf N}_{\rm st}^{\rm R}(\theta)
    = \eta(\theta){\hbar I\over 2e}~
    \hat{\bf m}\times(\hat{\bf m}\times\hat{\bf M}),
    \label{eqn:Nst}
\end{equation}
%%>>>>>>>>>>>>>>>>>>>>
where the cross product has magnitude of $\sin\theta$. In
Slonczewski's hybrid theory, \cite{Xiao:2004,Slonczewski:2002}
%%<<<<<<<<<<<<<<<<<<<<
\begin{equation}
    \eta(\theta) = {q_+\over A+B\cos\theta} + {q_-\over A-B\cos\theta}.
    \label{eqn:eta}
\end{equation}
%%>>>>>>>>>>>>>>>>>>>>
The parameters $A, B$, and $q_\pm$ are calculated using the material
parameters and geometries as shown in
Ref.~\cite{Xiao:2004}. An equivalent spin-transfer torque
formula was obtained by Manschot, et al., \cite{Manschot:2004a}
independently.

As discussed above in Sec.~\ref{sec:approx}, the drift diffusion
approximation assumes that the 
distribution function has a simple form consisting of a
uniform expansion and a contribution proportional to the velocity.
Thus, we can expect that the drift-diffusion approximation breaks down when the
variation of the distribution function over the Fermi surface is more
complicated. 
There are three situations 
where more complicated behavior is introduced.  The first is when the
transmission through the interface depends strongly on wave vector as
is typically the case \cite{Stiles:1996}.  In the immediate vicinity
of the interface, the wave-vector dependence of
the transmission gives the
distribution function a complicated variation over the Fermi surface.
This variation includes contributions that decay 
on the order of the mean free path, see \Eq{eqn:sol-homo-ex}.  If the
interfaces are separated by more than this length, the strong variation
decays between the interfaces, and the two approaches can be brought
into agreement through an appropriate choice of an effective interface
resistance in the drift diffusion approach.  However, when the
interfaces are closer, the interaction of these exponential
contributions between interfaces complicate the transport.  Evaluating
the importance of these effects requires a calculation using realistic
band structures, which is beyond the present calculations.  We instead
evaluate the other two situations where such difficult calculations
are not necessary.

\begin{figure}
    \centering
    \resizebox{0.9\columnwidth}{!}{%
    \includegraphics{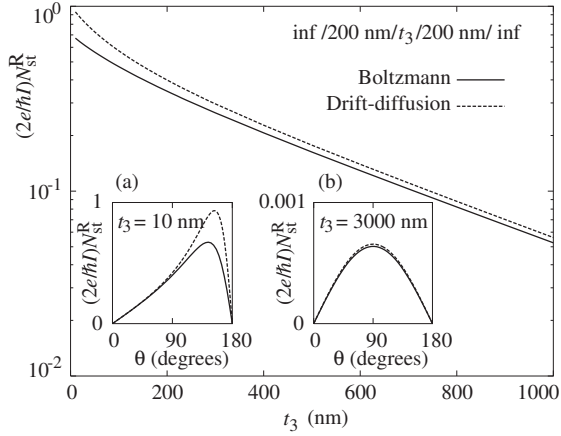}}
    \caption{Spin-transfer torque at the right interface of the spacer
    layer in a spin valve with semi-infinite leads.  Solid curves are
    calculated from the Boltzmann equation, dashed curves are from the
    drift-diffusion method.  The two insets show the angular
    dependence of the torque for two specific thicknesses,
    $t_3=$~10~nm (a) and 3000~nm (b).  The main
    panel shows the thickness dependence of the maximum values as a
    function of angle for the two approaches.  The legend gives the
    thicknesses of the layers (inf=infinite).}
    \label{fig:max}
\end{figure}

The second situation in which the distribution function has a
complicated angular dependence is when the spacer
layer is thin compared to its mean free path and the magnetizations
are not collinear.  \Fig{fig:max} compares the angular dependence of
the torque calculated with the drift diffusion approximation to
the torques calculated with the Boltzmann equation.  In these calculations,
the reflection at the interfaces has been set to zero so that the
complications described in the previous paragraph do not play a role.
Inset (b) in \Fig{fig:max} shows that the torques agree when the
spacer layer is very thick, and inset (a) shows that when they are
thin, there are significant differences.  The main panel shows the
variation of the maximum of the torque curves as a function of
thickness.  The difference between the curves gives the corrections
due to the complicated angular dependence of the distribution
function. 

The torques decrease with thickness for two reasons.  The large length
scale decay is set by the spin diffusion length.  When the layer is
thicker than its spin diffusion length, spin-flip scattering leads to a
significant decrease in the polarization of the current that crosses
from one side to the other.  For spacer layers thinner than their spin
diffusion length, but longer than their mean free paths, the
polarization of the current depends on the ratio of the effective
polarized resistance to the effective unpolarized resistance.  For
these structures, in which the ferromagnetic layers are thicker than
their spin diffusion length, the polarization of the current decays
roughly like one over the thickness of the spacer layer.

\begin{figure}
    \centering
    \resizebox{0.9\columnwidth}{!}{%
    \includegraphics{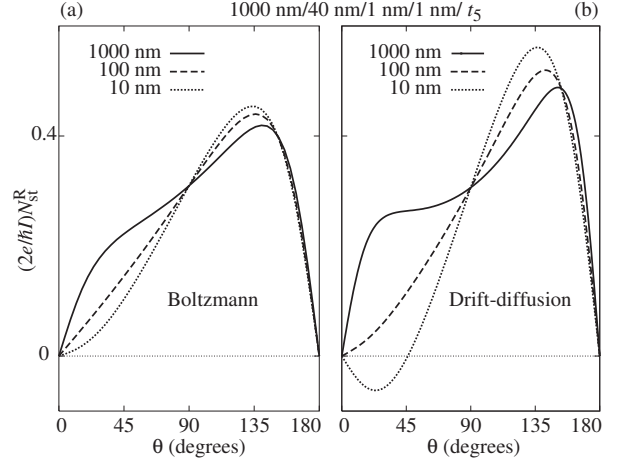}}
    \caption[Boltzmann vs. Drift-diffusion]{Spin-transfer torque at
    the right interface of the spacer layer in a spin valve. The left
    panel is calculated using the Boltzmann equation, the right panel
    using the drift-diffusion method.  The legend gives the
    thicknesses of the layers.}
    \label{fig:lead}
\end{figure}

The final situation, in which the drift-diffusion approach is not
adequate to describe the full angular dependence of the distribution
function, is at the interfaces between the leads and the reservoirs.
Typically, in the drift-diffusion approach, the spin accumulation is
set to zero at this point and the spin current is allowed to vary.
The argument is roughly that the large total density of states there
compared to in the leads forces the spin accumulation to be small.  In
Sec.~\ref{sec:finite_leads}, we described how the greater flexibility
available in the Boltzmann equation allows the implementation of
boundary conditions that treat the reservoir as an absorber.  In
\Fig{fig:lead}, we show the differences that can result from the
differences in the boundary conditions.  Both calculations show that
the angular variation in the torque depends strongly on the length of
the leads.  However, the Boltzmann equation results are not as
sensitive as those from the drift diffusion calculation.  In fact, the drift
diffusion calculation gives both the parallel and antiparallel states
as unstable for an asymmetric enough junction.  

\begin{figure}
    \centering
    \resizebox{0.9\columnwidth}{!}{%
    \includegraphics{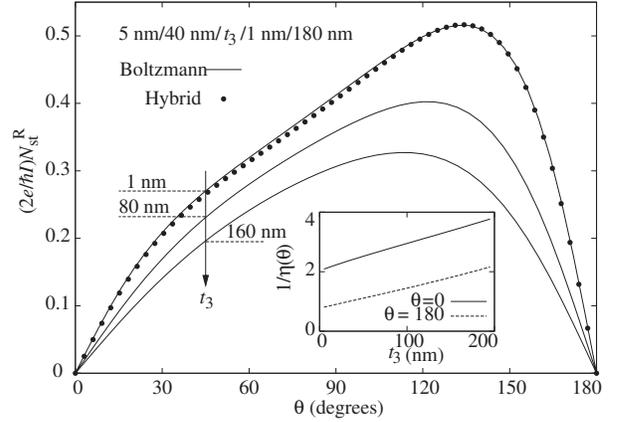}}
    \caption[Spin-transfer torque versus spacer layer length]{Spin-transfer
    torque at the right interface of the spacer layer in a spin valve with layer
    thicknesses 5~nm/40~nm/$t_3$/1~nm/180~nm with $t_3=1$ nm, 80 nm, and 160 nm.
    The solid curves are calculated from the Boltzmann equation. Solid circles
    are calculated by from the hybrid theory. The latter do not depend on $t_3$.
    The inset shows the $t_3$ dependence of $1/\eta(\theta)$ for
    $\theta = 0^\circ$ 
    and $180^\circ$ for this geometry.  The legend gives the
    thicknesses of the layers.}
    \label{fig:t3}
\end{figure}

The results described above show that the drift diffusion approach
does not work 
when the layers are thin.  Slonczewski
\cite{Xiao:2004,Slonczewski:2002}, developed a simple hybrid theory that
overcomes some of these difficulties.  In particular, it treats the
left going and right going electrons in the spacer layer separately.
This overcomes the errors illustrated in \Fig{fig:max}.  The theory
then treats the transport in the rest of the system with an approach
closely related to circuit theory \cite{Brataas:2000,Brataas:2001}.
The result is an analytic 
expression for the torque, \Eq{eqn:eta}.  Here, we compare this hybrid
theory with the Boltzmann equation to test its validity. In addition,
we explore the systematic behavior 
of the spin-transfer torque as a function of the spin valve
geometry. \Fig{fig:t3} shows the angular dependence of the
spin-transfer torque acting on the right FM (Co) layer for a spin
valve with geometry: \[ \rm
Cu(5~nm)/Co(40~nm)/Cu(t_3)/Co(1~nm)/Cu(180~nm). \] The spacer layer
thickness $t_3$ varies from 1 nm to 160 nm. The magnitude of the
spin-transfer torque reduces as 
spacer layer thickness $t_3$ increases.  
Features of the torque are discussed in Ref.~\cite{Xiao:2004}.

%%<<<<<<<<<<<<<<<<<<<<
%%>>>>>>>>>>>>>>>>>>>>

In Slonczewski's hybrid theory \cite{Xiao:2004,Slonczewski:2002},
scattering in the spacer layer is ignored. This means the spacer layer is
treated as a thin film. In the case $t_3=1$ nm, the spacer layer thickness
satisfies the condition of the hybrid theory. If we fit the spin-transfer
torque curve calculated from the Boltzmann equation using the spin-transfer
torque formula \Eq{eqn:Nst} from the hybrid theory (see \Fig{fig:t3} for the
fit), we find that the fitted interface resistance values agree with the
experimental values within 15 \%. This is very good agreement considering the
experimental values themselves are accurate only within 10 \% to 20
\%.  However, 
if the spacer layer thickness becomes comparable to the mean free path in Cu,
the torque curves (the solid curve in \Fig{fig:t3} with $t_3=80$ nm and 160 nm)
cannot be fit by the hybrid theory for any values of the interface
resistances.

The inset figure in \Fig{fig:t3} shows how $1/\eta(0^\circ)$ (solid line) and
$1/\eta(180^\circ)$ (dash line) vary with $t_3$ in the Boltzmann calculation. These
quantities are related to the critical current for initiating a magnetization
switching: from parallel (P) to antiparallel (AP) $J_{\rm P\ra
  AP}\propto 1/\eta(0^\circ)$ and from antiparallel to parallel
$J_{\rm AP\ra P}\propto 
1/\eta(180^\circ)$. So the curves in the inset figure of \Fig{fig:t3} also show that
the critical currents vary almost linearly with the spacer layer thickness
$t_3$, and both curves have similar slopes. 
Experimental measurements show the critical currents increasing with
spacer layer thickness \cite{Albert:2002}.

%%<<<<<<<<<<<<<<<<<<<<
\begin{figure}
    \centering
    \resizebox{0.9\columnwidth}{!}{%
    \includegraphics{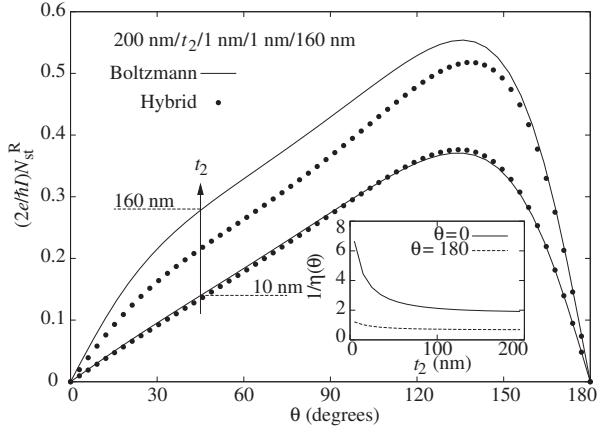}}
    \caption[Spin-transfer torque versus left ferromagnet length]{Spin-transfer
    torque at the right interface of the spacer layer in a spin valve with layer
    thicknesses 200~nm/$t_2$/1~nm/1~nm/160~nm with $t_2=10$ nm and 160 nm.
    Solid curves are calculated from the Boltzmann equation, solid circles are
    from the hybrid theory. The inset shows the $t_2$ dependence of
    $1/\eta(\theta)$ for $\theta = 0^\circ$ and $180^\circ$.  The legend gives the
    thicknesses of the layers.}
    \label{fig:t2}
\end{figure}
%%>>>>>>>>>>>>>>>>>>>>

%%<<<<<<<<<<<<<<<<<<<<
\begin{figure}
    \centering
    \resizebox{0.9\columnwidth}{!}{%
    \includegraphics{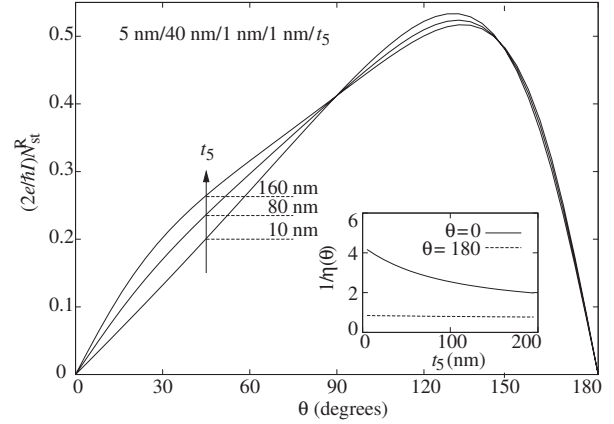}}
    \caption[Spin-transfer torque versus lead length]{Spin-transfer torque at
    the right interface of the spacer layer in a spin valve with layer
    thicknesses 5~nm/40~nm/1~nm/1~nm/$t_5$ with $t_5=10$ nm, 80 nm, and 160 nm.
    All solid curves are calculated from the Boltzmann equation. The inset shows
    the $t_5$ dependence of $1/\eta(\theta)$ for $\theta = 0^\circ$
    and $180^\circ$ 
    for this geometry.  The legend gives the
    thicknesses of the layers.}
    \label{fig:t5}
\end{figure}
%%>>>>>>>>>>>>>>>>>>>>

We have seen in \Fig{fig:t3} that Slonczewski's hybrid theory fails when the
spacer layer is thick. The breakdown of the hybrid theory is also seen in
\Fig{fig:t2}, where we show how the spin-transfer torque curve changes
with the thickness of the left ferromagnetic layer $t_2$.  The
input values in the hybrid theory here in \Fig{fig:t2} are the 
same as those used in \Fig{fig:t3}. In the case
$t_2=10$ nm, which is small compared to the spin flip length $l_{\rm sf}^F = 59$
nm in the ferromagnet, the hybrid theory and the Boltzmann calculation agree
with each other very well.  When $t_2=160$ nm, $t_2$ becomes comparable to or
larger than $l_{\rm sf}^F$, the hybrid theory starts to fail because
an approximation of the hybrid theory does not hold when 
$t_2 \gtrsim l_{\rm sf}^F$. This approximation assumes the spin currents at two
sides of the thick ferromagnetic layer are equal: $Q(x_1) \simeq Q(x_2)$ (see
\Fig{fig:scat-matrix}). But in this case of $t_2 \gtrsim l_{\rm sf}^F$, $Q(x_1)$
depends on $t_2$ in a non-trivial way.

Next, we study a spin valve with geometry: \[ \rm
Cu(5~nm)/Co(40~nm)/Cu(1~nm)/Co(1~nm)/Cu(t_5), \] where the right lead length
$t_5$ varies from 10 nm to 160 nm. \Fig{fig:t5} shows how the spin-transfer
torque curve acting on the second (thin) Co layer changes when we vary $t_5$. A
second bump around $\theta = 30^\circ$ appears in \Fig{fig:t5} as $t_5$ becomes
large. From the spin-transfer torque formula \Eq{eqn:Nst} and \Eq{eqn:eta} in
the hybrid theory, we see that the second bump corresponds to the $q_-$ term in
\Eq{eqn:eta}. The value of $q_-$ is typically close to zero and negligible, but
it becomes prominent when the spin valve is highly asymmetric.  By asymmetry, we
mean that the left and right sides of the spacer layer have different spin
dependent properties. For instance, for a spin valve with the geometry \[ \rm
Cu(5~nm)/Co(40~nm)/Cu(1~nm)/Co(1~nm)/Cu(160~nm),\] the left side of the spacer
layer has 5 nm Cu, and 40 nm Co, and two Cu/Co interfaces, which can be considered
mostly ferromagnetic, because both Co and Cu/Co interfaces have spin dependent
resistances. However, on the right side of the spacer layer, there is only 1 nm
of Co, while there are 160 nm Cu and two Cu/Co interfaces. So the 160 nm Cu
dilutes the ferromagnetic character of the Co bulk and the Cu/Co interfaces and
makes the right side of the spacer layer more like a non-magnet. This asymmetry
of the spin valve -- ferromagnet-like on the left and non-magnet-like on the
right -- leads to the emergence of the second bump in \Fig{fig:t5}.

%%----------------------------------------
\section{Summary} 
\label{sec:summary}

In summary, we developed a complete numerical algorithm to solve the
Boltzmann equation in multilayer heterostructures using a scattering
matrix formalism.  This method solves the spin-dependent Boltzmann
equation in a non-magnet and a ferromagnet and matches the bulk
solutions using an interface scattering matrix for the distribution
functions. The final solution for the distribution function is found
by imposing boundary conditions, either from infinite leads or from
the electron reservoirs. Our interest in using this method is to
calculate spin-transfer torque in a spin valve structure. The results
were found to agree with the Slonczewski's hybrid theory for
geometries typically encountered in experiments but not when layer
thicknesses become large compared to mean free paths. The drift-diffusion
method agrees poorly with the Boltzmann calculation due to the extreme
approximations it makes.

One of us (J.X.) is grateful for support from the Department of Energy
under Grant No. DE-FG02-04ER46170.

\end{document}